# A study of the wiggle morphology of HH 211 through numerical simulations

Anthony Moraghan [1][*], Chin-Fei Lee [1], Po-Sheng Huang [1], & Bhargav Vaidya [2]
[1] *Academia Sinica Institute of Astronomy and Astrophysics, P.O. Box 23-141, Taipei 106, Taiwan*
[2] *Department of Physics, University of Torino, Italy*

**ABSTRACT**

Recent high-resolution high-sensitivity observations of protostellar jets have shown many to possess deviations to their trajectories. HH 211 is one such example where sub-mm observations with the SMA have revealed a clear reflection-symmetric wiggle. The most likely explanation is that the HH 211 jet source could be moving as part of a protobinary system. Here we test this assumption by simulating HH 211 through 3D hydrodynamic jet propagation simulations using the PLUTO code with a molecular chemistry and cooling module, and initial conditions based on an analytical model derived from SMA observations. Our results show the reflection-symmetric wiggle can be recreated through the assumption of a jet source perturbed by binary motion at its base, and that a regular sinusoidal velocity variation in the jet beam can be close to matching the observed knot pattern. However, a more complex model with either additional heating from the protostar, or a shorter period velocity pulsation may be required to account for enhanced emission near the source, and weaker knot emission downstream. Position velocity diagrams along the pulsed jet beam show a complex structure with detectable signatures of knots and show caution must be exercised when interpreting radial velocity profiles through observations. Finally, we make predictions for future HH 211 observations with ALMA.

## 1 INTRODUCTION

Protostellar outflows often appear as a train of bright emission knots terminating in a leading bow-shock. One theory to explain the knots is that of episodic accretion and ejection events from the protostellar system that leads to a jet beam of variable velocity (e.g. Fendt 2009). Due to the small scales involved, as well as heavy obscuration around the protostellar sources, it is difficult to directly observe the precise mechanism behind jet launching and collimation. However, by studying the properties and morphology of the jets and outflows on extended scales, and with the help of models, we can extrapolate back and deduce the physical processes occurring near the protostellar source.

With the advent of modern high-resolution high-sensitivity instruments, misalignment or wiggles along outflow trajectories have been detected in several sources. For example, HH 111 shows two types of bending; a reflection-symmetric (or mirror-symmetric) wiggle within 2′ from the source (Noriega-Crespo et al. 2011), and point-symmetric wiggle at greater distances (Reipurth et al. 1997). HH 30 shows a C-shaped bending (Estalella et al. 2012). Similarly, L1448N shows a C-shaped bending and a wiggle of period ∼15-20 years (Hirano et al. 2010). HH 211 shows precession, C-shaped bending, and a distinct reflection symmetric wiggle (Lee et al. 2010). A jet propagating from a star in a binary system would develop a reflection-symmetric wiggle due to the orbital motion of the jet source, whereas a precession of the jet axis would produce a point-symmetric wiggle shaped jet (Raga et al. 2009). In this paper we investigate the reflection symmetry of HH 211 in more detail using numerical simulations.

The HH 211 protostellar outflow was first discovered through shock excited $H_2$ gas in the near-infrared by McCaughrean et al. (1994). It is one of the youngest outflows known thus far having a kinematical age of ∼1000 yr, and consists of two bright bow shocks of 0.13 parsec (pc) separation with a chain of knots in between (Dionatos et al.

---

[*] E-mail: ajm@asiaa.sinica.edu.tw

2010). HH 211 has been previously observed with the Submillimeter Array (SMA) (Ho et al. 2004) and the data extensively studied, including follow up observations, by Lee et al. (2007, 2009, 2010, 2014). These studies have resolved two sources in the central region; SMM1 and SMM2 with a separation of ∼84 AU. SMM1 is believed to be the driving source of the HH 211 jets and may itself be an unresolved protobinary as suggested by the reflection symmetric morphology of the outflow (Lee et al. 2010). The mass ratio of the two SMM1 components has yet to be determined, but their combined mass is estimated to have a lower limit of ∼50 $M_{Jup}$ (Lee et al. 2010). The SMM2 core is of planetary mass and may have broken away from the primary core (Lee et al. 2010). It is interesting that in recent years protostellar multiplicity has been found to be commonplace (e.g. Pineda et al. 2015), requiring an additional layer of complexity to the traditional models of star formation.

The velocity of the HH 211 jets have been measured as 170±60 km s$^{-1}$ from their proper motion and their trajectory has a slight inclination angle of ∼5° to our line of sight (Lee et al. 2009). More recently Jhan & Lee (2015) used a multi epoch study and revised the velocity to 115±50 km s$^{-1}$ and the inclination angle to ∼9°. The outflow structure consists of a series of symmetric knots within ±17″ of the driving source designated as BK1-6 on the blueshifted side, and RK1-5 plus RK7 on the redshifted side. Beyond this, peaks of H$_2$ $v$=0-0 S(5) emission designated as B1-2 and R1-2 appear to trace the outer bowshock wings where there are lower excitation conditions (Dionatos et al. 2010).

The HH 211 outflow is effected by different misalignment mechanisms. Early observations at lower resolution, such as McCaughrean et al. (1994) at 3″ resolution and Gueth & Guilloteau (1999) at 1.5″, resolution detected some precession and bending in the highly collimated jet structure. At 0.3″ resolution and the higher-sensitivity of the SMA, an additional misalignment component in the form of a reflection-symmetric wiggle was detected by Lee et al. (2010). These authors proposed it was due to the motion of the jet source within a binary system and derived an analytical fit to determine the orbital parameters. By studying the observed wiggle structure in terms of amplitude and periodicity, we can place constraints on the properties of the binary and the accretion/ejection processes occurring at the source.

The usual way to link observation with theory is through numerical simulations with hydrodynamic codes. Protostellar jets were first investigated through numerical simulations ∼30 years ago (e.g. Norman & Winkler 1985; Norman et al. 1982). Since that time, computer power has increased exponentially, and hydrodynamic codes have become extremely advanced including complex physics such as molecular chemistry and magnetohydrodynamics. Here we focus on the larger scale and employ the PLUTO hydrodynamics code (Mignone et al. 2007) to perform jet propagation simulations of HH 211 using initial conditions and the analytical model derived from the observations of Lee et al. (2010) in an attempt to reproduce the reflection symmetric wiggle, and make direct comparisons to the observation.

The layout of the paper is as follows; We describe our computational code and model in § 2. The results are discussed and compared to the observations using synthetic observations in § 3, and we conclude in § 4.

## 2 NUMERICAL METHODS

Jet propagation simulations of the HH 211 bipolar outflow are performed using the PLUTO V4.1 astrophysical computational code (Mignone et al. 2007). PLUTO is a modern multi-physics, high-resolution, high-order shock-capturing Godunov code. The code is very adaptable to tackle a diverse range of astrophysical problems. It was developed with a modular design and includes a variety of physics and solver algorithms, including the Adaptive Mesh Refinement (AMR) technique (Mignone et al. 2012).

PLUTO V4.1 introduced a new molecular cooling and chemistry module. It is in the form of a full time-dependent cooling function that self-consistently tracks the destruction and formation of molecular hydrogen, atomic hydrogen, and ionized hydrogen. For greater accuracy, the module also employs an equation of state that takes into account the degrees of freedom arising from the rotational and vibrational modes of H$_2$. See Vaidya et al. (2015) for a more detailed description and comprehensive tests of the cooling module.

The number fraction of the different hydrogen species mentioned above are obtained using the cooling module. In the present work, we focus on comparing synthetic CO maps obtained from simulations with those obtained from SMA observations of HH 211. The fraction of CO required for synthetic maps is estimated by converting the H$_2$ fraction under a simplified assumption of a constant CO/H$_2$ abundance ratio. Explicit treatment of CO chemistry (e.g. Raga et al. 1995; Rosen & Smith 2003) and its micro-physical effects on the equation of state will be incorporated in future works.

We should note that unlike the jet launching simulations from a binary system performed by Sheikhnezami & Fendt (2015), we only perform jet propagation simulations starting close to the protostar by assuming that the jet ejection and collimation process has taken place just outside the computational domain. Although the base of the jet should be close to the protostar, we do not include any gravitational or heating effects. We simply mimic the effect of binary rotation by perturbing the base of the pre-collimated jet beam.

### 2.1 Initial conditions

In order to reduce the amount of computational time and resources required for the simulation, we model only one jet, and then mirror the result in post-processing to create the full reflection symmetric bipolar system. The jet beam is introduced through the lower boundary of the 3D computational domain (the $x-y$ plane) as a circular pre-collimated beam of 20 AU in radius. The total length of the computational domain ($z$-direction) is set as 6000 AU and the transverse distance on each side of the jet beam to the boundary is 225 AU. Therefore the total 3D domain represents a volume of 6000×450×450 AU. The lower boundary has reflective boundary conditions, except inflow boundary conditions at the jet beam inlet. All other boundaries have outflow boundary conditions.

In order to perform the simulations at as high a resolution as feasible and as computationally efficient as possible, we implement AMR. We apply a base 'level 0' grid of 480×36×36 zones and allow for 2 levels of refinement. There-fore the effective resolution at 'level 2' is 1920×144×144 zones and 3.125 AU per zone.

The orbital motion is considered from the analytical description of the jet beam wiggle as described by Lee et al. (2010). It was derived to fit the observed wiggle in the SMA images of HH 211. In the analytical model, the jet source is simply moving in a circular orbit of radius 2.3 AU with a period of 43 years (see Lee et al. 2010, section 3). With this assumption, the trajectory of the model is seen to accurately trace the observed jet pattern. The amplitude of the wiggle, $x$, with distance, $z$, along the axis is defined by Eq. 6 from Lee et al. (2010) as

$$x \approx -\kappa z \cos\left(\phi_0 - \frac{2\pi z}{\Delta z} + \frac{\pi}{2}\right) + \eta z \quad (1)$$

where $\kappa = 0.0094$ radians (0.54°) is the velocity ratio between the orbital velocity, $v_0$, and the jet velocity $v_j$, $\phi_0 = 2.705$ radians (155°) is the phase angle, $\Delta z$=1530 AU is the wiggle period, and $\eta = 0.0096$ radians (0.56°) is the additional C-shaped bending component to the north.

To simulate the analytical model, we move the jet source in a circular orbit of 2.3 AU with the same orbital period of 43 years along the grid boundary. The position of each zone inside the jet beam varies by a maximum of 4.6 AU during an orbit. As the effective resolution of the grid is 3.125 AU, the physical orbital motion only barely resolved. However, the centrifugal velocity components from the orbital motion are the main factor leading to the wiggle morphology (e.g Masciadri & Raga 2002), and those components of velocity are included in the simulation through the jet beam area that covers ~128 zones at the jet inlet.

Recent Spitzer observations of HH 211 by Tappe et al. (2012) have estimated molecular gas densities of $10^6$–$10^9$ cm$^{-3}$ in regions associated with CO band emission. This is in agreement with SMA observations of Lee et al. (2009, 2010). Therefore, we apply an ambient gas density of $1\times10^6$cm$^3$ at our jet source and include an ambient density profile with distance, $r$, from the source as $(1/(1+r/r_0))^2$, where $r_0$ is set as 2000 AU to represent a uniform core region. We also apply a corresponding pressure profile so that the entire ambient medium is at an initial temperature of 10K. The jet beam is introduced as over-dense with density $2.8\times10^6$cm$^3$ and a temperature of 100K. Both the jet beam and ambient medium are considered fully molecular.

As well as a simulation of a steady velocity jet with orbital motion, we attempt to reproduce the visible knotty structure of HH 211 by including a sinusoidal velocity variation to the jet beam. When high velocity material catches up with slower velocity material a knot forms. Lee et al. (2007) were first to propose that the observed knots may be a consequence of perturbations caused by the presence of an unresolved binary a few AU from the jet source. They also measured the velocity range of the knots as ~25 km s$^{-1}$. From the observational results of Lee et al. (2010), the average jet beam velocity is 170 km s$^{-1}$ and there are 7 knots visible within 15″ (5000 AU) of the source. This means a new knot has formed every 20 years on average which is roughly half of the predicted 43 year orbital period from the analytical model. With these values in mind, we set the base jet beam velocity in our simulation as 170 km s$^{-1}$ and include a sinusoidal velocity variation at half the orbital period, 21.5 years. In one model the velocity variation is set as ±20 km s$^{-1}$, and in the other model it is set as ±50 km s$^{-1}$.

We conserve momentum during the velocity pulsations. When the velocity is in the decreasing phase of the pulsation, the density and pressure are increased by an equivalent amount to conserve momentum. Using the jet beam density, jet radius, and base jet velocity, we calculate the average mass loss rate of our simulated bipolar outflow to be ~1.4×10$^{-6}$M$_\odot$yr$^{-1}$. This value is ~1.28 times lower than the observed mass loss rate of 1.8×10$^{-6}$M$_\odot$yr$^{-1}$ as estimated by Lee et al. (2007), and ~1.16 higher than the 1.2×10$^{-6}$M$_\odot$yr$^{-1}$ value estimated by Jhan & Lee (2015).

### 2.2 Radiative transfer routine

Gas density and molecular hydrogen fraction, for example, cannot be directly observed. However, we can use the hydrodynamic quantities to post-process the simulated data into a synthetic observation in order to directly compare with observations. Therefore, we use a radiative transfer routine to create synthetic CO $J$=3-2 emission maps for direct comparison to CO $J$=3-2 transitions at 345 GHz as observed with the SMA sub-mm array. The procedure is as follows;

The one-sided jet from the PLUTO code is turned into a bipolar flow by reflecting the data on the $x-y$ plane and changing only the jet propagation velocity component (V2) on the reflected side. This reflection symmetric bipolar outflow is then inclined by 5° to our line of sight to match the observed inclination angle of HH 211. Assuming a distance of 280 pc to HH 211 and the length of the computational domain as 2×6000 = 12,000 AU, the total size of this 3D computational domain is considered to be ~40″×1.6″×1.6″. Our radiative transfer code uses the density, temperature, velocity, and molecular fraction data to create a synthetic CO emission channel map covering a velocity range of ~-30–40 km s$^{-1}$ that can be convolved for any instrumental resolution and sensitivity. As we do not yet track CO abundance directly in the PLUTO molecular chemistry module, we estimate it from the H$_2$ abundance using the commonly accepted conversion factor of 4×10$^{-4}$. Finally a slight bending 0.56° to the north is applied to each bipolar component as observed in HH 211 (Lee et al. 2010).

Although the actual HH 211 outflow has propagated to distances of ±15000 AU (McCaughrean et al. 1994), we are only interested in the inner region of ±17″ (±5000 AU) that shows the distinctive internal knots and was studied through the Lee et al. (2010) SMA observation. Our simulated domain covers ±20″ (±6000 AU) and we continued the simulation until the leading bowshock has left the grid in both the steady and pulsed velocity cases. However, we let the pulsed velocity simulation propagate much further off the grid in order to ensure there is no bowshock material present on the grid as it could cause additional synthetic emission.

## 3 RESULTS AND DISCUSSION

The basic hydrodynamic simulation results are shown in Figure 1 depicting a 2D slice through the 3D domain showing log of density for both the steady (panel a) and the 170±20 km s$^{-1}$ velocity (panel b) simulations, followed by velocity (panel c), log of temperature (panel d), and H$_2$ fraction (panel e) for the 170±20 km s$^{-1}$ velocity simulation.

In the density slice of the steady velocity jet (panel a), we see the orbital motion of the source leads to a clear wiggle-shaped trajectory of the jet beam. It can be compared directly to the density slice of the pulsed velocity jet (panel b). Here we see the combination of the wiggle motion and velocity pulsation leads to progressively more oblique internal shocks. We illustrate the orientation of the oblique shocks in the 2D plane by the blue lines drawn on the lower H$_2$ fraction plot (panel e). The displayed evolution time of the steady jet is ∼233 years, and that of the pulsed jet is ∼320 years. Both times are at a phase angle of 155° in the orbital motion of the jet source as fitted in the analytical model of Lee et al. (2010). The pulsed jet simulation has undergone another two complete orbits of ∼43 years each in order to ensure the leading bowshock has propagated well outside the ±17″ region of interest. In both the steady and pulsed velocity cases, the outer outflow boundaries have also left the grid. Here we are only interested in the jet beam and use of a larger domain would have been more computationally demanding.

In the velocity and temperature plots (panels c and d) we can detect a developing shock at a distance of 600AU from the source. Beyond which we see a series of seven equally spaced fully developed internal shocks or 'knots', of increasing obliquity to the jet propagation axis. The sinusoidal velocity variation of the jet beam at ±20 km s$^{-1}$ creates the internal shocks. In the density plot in particular, the shape of the shock front for each internal shock shows trajectories increasingly deviating from the central jet axis with distance from the source. In three dimensions the shocks resemble skewed arc-like structures. The orbital motion at the base of the jet beam imparts centrifugal velocity components that increase in amplitude with distance from the source leading to the wiggle-shaped trajectory. Thus, while the high velocity jet material is shocking against the slower moving material, the position of the Mach disk (the point where the jet beam is decelerated at the shock front) is slowly changing position. Then, when viewed perpendicular to the jet beam propagation direction, the shocks appear at oblique angles. Comparing the density scales of the steady and pulsed velocity simulations, the pulsed velocity case has much higher localised regions of density (∼12 times higher) occurring at the shock fronts. These regions should lead to strong synthetic emission.

The centrifugal velocity of the orbital motion at 1.6 km s$^{-1}$ causes the jet beam to widen downstream into a cone-like volume with opening angle, $\kappa = 0.54°$. At a distance of 4500 AU from the source, the jet beam of the steady velocity simulation has expanded to an effective radius of 56 AU. Similarly, by the time the pulsed velocity jet beam reaches the end of the grid (at a distance of 6000 AU), the jet beam radius has increased by almost 4 times from its initial 20 AU radius to an effective radius of 76 AU.

In the temperature plot, along the jet beam, we see temperature jumps from 100K to ∼1000K across the knots. The interface between the jet beam and the cavity possesses a much higher temperature of ∼10,000K, especially near the jet beam inlet where higher ambient densities are present.

The entire cocoon remains very well collimated as expected for the case of an over-dense jet associated with protostellar environments. The sharp V-shaped leading bowshock is a consequence of the molecular cooling processes and is apparent in other 2D and 3D simulations of jets with cooling (e.g. Moraghan et al. 2008; Teşileanu et al. 2014) where the cooling at the bowshock creates a low pressure working surface leading to a reduced rate of material ejected sideways into the cocoon.

In the molecular hydrogen plot we see the undisturbed ambient medium and jet beam are fully molecular. The ±20 km s$^{-1}$ velocity variation in the jet beam does not produce shocks strong enough to dissociate the molecular gas. However, molecular gas is dissociated in the cavity after passage through the leading bowshock. In the low-density hot gas of the cavity, reformation times of molecular hydrogen are long. But as we look closer towards the base of the jet, the fraction of molecular hydrogen increases where there is denser and older shocked gas in an environment of higher pressure.

### 3.1 Synthetic CO integrated intensity maps

In Figure 2 we directly compare the region of ±17″ (±5000 AU) from our steady and pulsed velocity jets to the SMA observational CO emission map of HH 211 from Fig 1b. of Lee et al. (2010). The observational map is convolved to 0.″46 × 0.″36 resolution (top panel) and we convolve the simulated maps to the same resolution for direct comparison (steady velocity jet case in the second panel, and pulsed velocity jet cases in the lower two panels).

The analytical model of Lee et al. (2010) is traced by the red and blue lines in the observational map, and the green lines in the simulated maps. The steady velocity jet in particular shows very good agreement by precisely matching the wiggle trajectory. This is to be expected as the initial conditions of the simulation were based on the analytical model parameters. It shows that the observed reflection symmetric wiggle of HH 211 could indeed be explained by the jet source as being part of a protobinary system. Similarly, in the pulsed velocity jet, we see that the emission knots also closely trace the wiggle pattern. However, as we shall see, a more complex model involving a different pulsation pattern for the red and blueshifted sides may be needed in order to account for the precise pattern of knots along the outflow.

The CO contour levels of the simulated maps are plotted at the same intensities as the observational plot, beginning at 0.56 Jy beam$^{-1}$ km s$^{-1}$ and stepping at intervals of 0.84 Jy beam$^{-1}$ km s$^{-1}$, and covering the same radial velocity range; 20 to 43 km s$^{-1}$ and -16 to 0 km s$^{-1}$, for the red and blue shifted components, respectively. For the steady velocity jet at 170 km s$^{-1}$, we only see the base contour at 0.56 Jy beam$^{-1}$ km s$^{-1}$ smoothly tracing the jet beam material.

Two pulsed velocity jets are presented; one with a velocity of 170±20 km s$^{-1}$, and one with a slightly stronger perturbation amplitude of velocity 170±50 km s$^{-1}$. In both cases the pulsation period is the same, 21.5 years [1]. The ±50 km s$^{-1}$ case emphasises that the no matter what the velocity perturbation amplitude is, the wiggles are consistent with observation. The 21.5 year pulsation period was chosen as it can create the correct number of HH 211 knots within the observed distance. However, there are some discrepancies when we match the positions of the knots between the observation and simulation. We label the simulated knots as SBK1–6 on the blueshifted side and SRK1–6 on the redshifted side, and compare their positions directly to the observational knots labelled by Lee et al. (2010) as BK1–6 and RK1-5, and 7 (top panel).

---

[1] Although the displayed times in all cases are at a phase angle of 155°, the ±50 km s$^{-1}$ case is presented at an earlier stage of evolution two orbital periods behind the ±20 km s$^{-1}$ case, or ∼86 years earlier. It is at the same evolution time as the steady jet simulation.

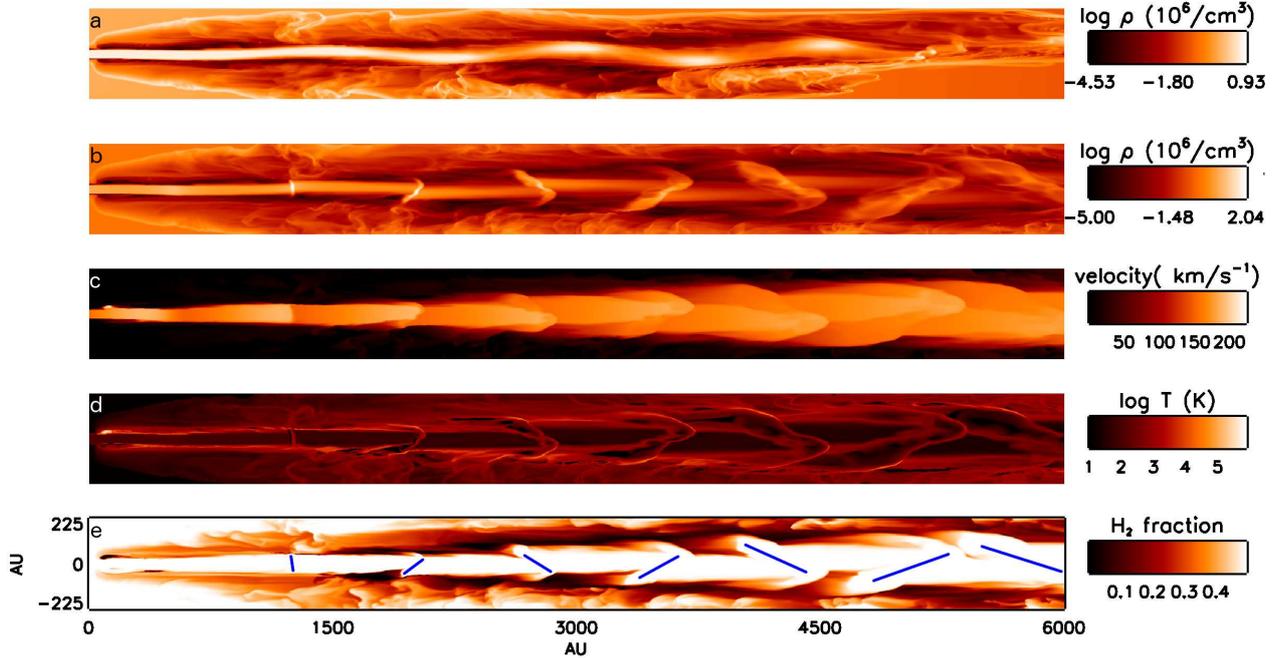

**Figure 1.** 2D slices through the center of 3D domain of our simulations. Panel (a) shows log of density of our steady jet velocity simulation. It can be directly compared to (b) showing log of density of the 170±20 km s$^{-1}$ velocity simulation. The remaining panels show the velocity (c), log of temperature (d), and H$_2$ fraction (e), of the 170±20 km s$^{-1}$ velocity simulation. The wiggle-shaped trajectory from the orbital motion, combined with the jet beam velocity pulsations lead to progressively angled internal or 'oblique' shocks. Blue lines in the H$_2$ fraction plot are drawn to illustrate the orientation of these oblique shocks in the displayed 2D plane.

### 3.2 Knot positions

On the redshifted side, the alignment of the observational and simulated knots are quite good. RK2, RK3, RK4, RK5, and RK7 are mostly coincident with SRK2, SRK3, SRK4, SRK5, and SRK6. Only SRK4 and SRK5 are slightly ahead of RK4 and RK5 by ∼1″. Note the positions of the observational knots were determined from SiO intensity peaks, and the positions of the simulated knots are determined from the peaks of CO intensity. Although there could be slight position shifts between CO and SiO (Lee et al. 2015), here CO could trace weaker (postshock) shocks.

In contrast, there are poorer alignments on the blueshifted side. The SBK2 knot lags behind BK2 by ∼1″, whereas SBK3 and SBK4 are ahead of BK3 and BK4 by 1″ and 1.5″, respectively. SBK5 does not align well with BK5 but aligns better with BK6. The unlabelled emission at 15″ in the observational image aligns well with SBK6.

This comparison shows there is some merit to the 21.5 year pulsation period, especially for the redshifted side. The major knots are reproduced, but the blueshifted side appears more complex. It suggests that knots may not be produced simultaneously on the red and blueshifted sides of the protostar. Although the positions of BK1/RK1 are equidistant from the source, the positions of the later knots do not have equidistant positions on both sides. This was also noted in a recent multi-epoch observational study of HH 211 by Jhan & Lee (2015). It is common to measure different flow velocities on each side of the bipolar systems (Correia et al. 2009). Theory even proposes that one sided jets are possible (e.g. Dyda et al. 2015), and have been observed (e.g. Gómez et al. 2013). Different outflow mass fluxes were noted between the upper and lower hemispheres of the recent 3D MHD jet launching simulations from a binary system by Sheikhnezami & Fendt (2015). So ejection velocities and periods may be different on each side of the bipolar system and not completely regular and symmetric as assumed in our model.

The CO contour levels in the observational plot begin at 0.56 Jy beam$^{-1}$ km s$^{-1}$ and step at intervals of 0.84 Jy beam$^{-1}$ km s$^{-1}$. Again we plot the simulated emission map with the same contour levels. The observational contours show high emission near the source, and weaker emission further downstream at the BK6 and RK7 knots. In contrast, the contours on the simulated plot show the opposite effect; almost no emission near the source, and stronger emission in the downstream knots.

The amplitude of the velocity variation is not well constrained from the observations and the velocity jumps across internal working surfaces rapidly decrease with distance from the source (Raga & Kofman 1992). That is why we investigate stronger amplitude of perturbation of the ±50 km s$^{-1}$ case. The ±50 km s$^{-1}$ case shows stronger knot and inter-knot emission in the outer bow-shocks compared to the ±20 km s$^{-1}$ case. However, as the ±50 km s$^{-1}$ case is presented two rotation periods (∼86 years), earlier than the ±20 km s$^{-1}$ case, the stronger emission may be the result of being closer to the leading bowshock and not yet propagating through such an excavated cavity as would be the case during later times of the evolution.

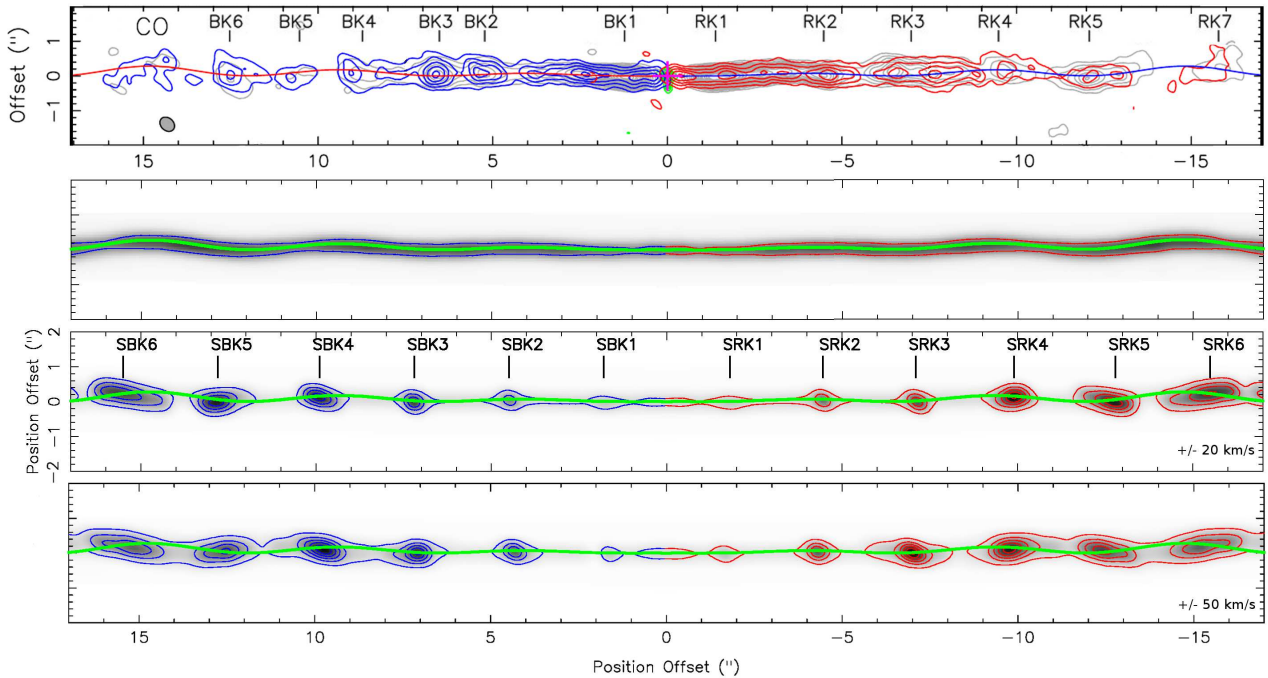

**Figure 2.** Top panel: A reproduction of Fig. 1b from Lee et al. (2010) showing the CO emission integrated intensity map of HH 211 within 17″ from the source. The red and blue contours highlight the red and blue shifted components integrated from 20 to 43 km s$^{-1}$ and -16 to 0 km s$^{-1}$, respectively, where the contours begin at 0.56 Jy beam$^{-1}$ km s$^{-1}$ and step at intervals of 0.84 Jy beam$^{-1}$ km s$^{-1}$. The location of the blue and red-shifted knots are labelled. The red and blue lines trace the wiggle produced by the analytical model. Second panel: Simulated synthetic CO emission map for the steady velocity jet simulation. For a direct comparison to the observation, it uses the same channel map velocity range and contour intensity values. The green line traces the analytical model wiggle. This shows that the assumption of the jet source in a binary system can indeed reproduce the reflection symmetric wiggle. Third panel: Simulated synthetic CO emission map of the 170±20 km s$^{-1}$ pulsed jet simulation. Again the channel map velocity range and contour intensity values are the same as the observational image and the green line traces the analytical model wiggle. We see that whereas the observational map has intense jet emission near the jet source, the simulated map has weaker emission near the source. Also, the positions of the knots are not precisely aligned with the positions of the observational knots, especially in the blueshifted side. However, the intensity range of the simulated knots are similar to the observational knots. Fourth panel: Similar to the third panel, but for the 170±50 km s$^{-1}$ pulsed jet simulation. Note the displayed time is at an earlier phase of evolution. It is two rotation periods before that of the 170±20 km s$^{-1}$ jet in the third panel. This may account for increase of intensity of the outer knots as they are closer to the leading bowshock. However, it shows a high amplitude of velocity perturbation can not increase the intensity of emission near the source in our model.

Now we make a direct comparison between the ±20 km s$^{-1}$ and ±50 km s$^{-1}$ cases, this time comparing them at an earlier, but identical phase of evolution (156 years), and zoomed into the red-shifted region between 0″ to -9″ near the source. The resulting integrated emission map is shown in Figure 3. Here we cover the entire radial velocity range so that the contours now highlight the slow moving cocoon material as well as the jet material. The displayed contour levels are the same as before; beginning at 0.56 Jy beam$^{-1}$ km s$^{-1}$ and stepping at intervals of 0.84 Jy beam$^{-1}$ km s$^{-1}$. There is little morphological difference between the two cases, except the higher velocity perturbation creates broader shocks. This is because the stronger shock velocity heats the gas more, enabling it to expand into a larger volume, which in turn allows it to cool more efficiently. In terms of synthetic emission, this is apparent as slightly stronger emission from one extra set of contours.

Our model cannot account for the innermost knots, BK1 and RK1. The simulated innermost knots SBK1 and SRK1 are about ∼1″ ahead of BK1 and RK1 although they have not fully formed as strong shocks. The stronger velocity perturbation of the ±50 km s$^{-1}$ does not help either. In reality, the presence of a protostar and envelope may introduce higher excitation temperatures and densities near the source leading to enhanced emission. Our model does not contain any additional forms of heating. Only a pre-collimated beam with a temperature of 100K is introduced at the source. Another possibility could be a series of velocity pulsations of shorter period creating a chain of closely spaced sub-knots that then appear as the band of continuous emission in the observation. The sub-knots may eventually converge into the larger knots seen downstream (e.g. Yirak et al. 2009; Raga et al. 2012). This was postulated by Lee et al. (2010) to account for wide velocity range detected in the Position-Velocity diagrams near the source (see Section 3.4). Of course a full accretion-ejection model for a binary system such as Sheikhnezami & Fendt (2015) would be required to accurately model the region near the jet source. However, our model is mainly focussing on the large scale morphology of the outflow. It is computationally challenging to perform full scale self-consistent simulation containing an MHD accretion-ejection mechanism and a propagation region of several thousand AU in length.

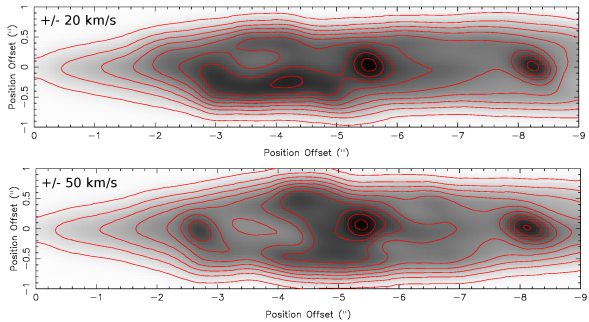

**Figure 3.** Integrated CO emission intensity maps of the 170±20 km s$^{-1}$ and 170±50 km s$^{-1}$ velocity perturbation cases. Here we focus only on the region of the red-shifted side from position offset 0″ to -9″, and integrated over the entire radial velocity range in order to show emission from both the cocoon and the jet. The displayed evolution time is at a time of 156 years, earlier than the times shown in the previous plots of this paper. The contour levels are as before; beginning at 0.56 Jy beam$^{-1}$ km s$^{-1}$ and stepping at intervals of 0.84 Jy beam$^{-1}$ km s$^{-1}$. The faster velocity perturbation leads to slightly broader shocks, and stronger emission evident from an extra level of contours.

### 3.3 21.5 year sinusoidal velocity variation

Our model has merit as it can account for the correct number of major knots downstream in roughly the same locations. It is well established that a magneto-hydrodynamic process controls the accretion mechanism through the transfer of material from the accretion disk to the protostar. In order to shed excess angular momentum, some material is expelled away from the system by the magnetic fields to become the jet. However, the precise mechanism behind the accretion and ejection process remains uncertain.

Physically, a knot forms from a supersonic velocity jump when significant variations of ejection speed or angle leads to new faster material catching up with slower and older ejecta (e.g. Völker et al. 1999). For HH 211, the number of major knots compared to the orbital period suggest there are two perturbations per orbit. Assuming the HH 211 source is part of a close binary system, it is reasonable to assume that a close encounter to either the companion or the circumbinary disk may provoke a variation of the accretion/ejection mechanism and hence affect the jet velocity. Interestingly it has been noted that a misaligned protostellar disk in a binary system wobbles with a period of approximately half the orbital period of the binary (Bate et al. 2000). If this is the case, we could assume that feedback from the wobble leads to an enhanced ejection event.

We also briefly speculate on other possibilities relating to variable jets from sources in elliptical orbits. If the system has a large eccentricity, a perturbation of the jet source accretion disk may occur at the periastron passage to the companion protostar (e.g. Garcia et al. 2013). This was noted in observations of a spectroscopic binary protoplanetary nebula by Witt et al. (2009). These authors noted maximum accretion speeds occured near periastron, and minimum accretion speeds occured near apoastron during the highly eccentric ($e = 0.37$) orbit.

But to account for the separation distances between the HH 211 knots, it would require two perturbations per orbit, therefore we could assume another perturbation may occur at the apoastron of the orbit when there would be a close approach to the inner edge of a circumbinary disk, or just a relaxation effect after the periastron perturbation. However, the orbit must be eccentric for this to be the case, and in an eccentric orbit the time spent during the apoastron passage is longer than the periastron passage. This could effect the mass and sizes of the observed knots where compact knots of lower mass and momentum occur during the fast periastron passage, and more extended knots occurs during the longer apoastron passage (González & Raga 2004). Similarly, another possibility could be perturbations from two approaches to the inner edge of the circumbinary disk during the orbit, again if the orbit is eccentric.

The recent full 3D MHD binary jet launching simulations of Sheikhnezami & Fendt (2015) have revealed some interesting effects. They found sudden increases in mass flux are triggered by the misalignments of the disk due to tidal effects of the companion star, and that the outflow mass fluxes differ between the upper and lower hemispheres of their model. It was noted through simulations of time-dependent ballistic jets that there may be no major differences when comparing the trajectory of a jet from an orbiting source to a precessing jet at large distances from the source (Masciadri & Raga 2002), nor major differences of the jet trajectory between a circular or elliptical orbit (Velázquez et al. 2013). Therefore studying discontinuities (knots) in the jet flow, rather than the jet trajectory, may be easier in determining the eccentricity of the orbit of a jet source.

A more detailed observational study of the knots at higher resolution and higher sensitivity is required to accurately compare the sizes and properties of the knots. In this paper, our simulation produces knots of equal size through a regular sinusoidal perturbation. *Our results show that a regular sinusoidal velocity perturbation can reproduce the correct number of observed knots in HH 211 downstream, but cannot precisely account for the observed knot positioning on the blueshifted side, nor match the strength of the innermost knots.*

### 3.4 Position-Velocity diagrams

In Figure 4 we plot a position-velocity (PV) diagram along the length of the jet beam for both our steady and 170±20 km s$^{-1}$ pulsed velocity cases. The PV diagrams correspond to a wide-slit that includes the entire cross section of the observed jet. We only show the redshifted side and the velocity range from -15 – 9 km s$^{-1}$. This time the displayed contour levels begin at 0.1 Jy beam$^{-1}$ and step in intervals of 0.1 Jy beam$^{-1}$ in order to highlight the jet beam material. As expected, there is a distinct difference between the steady and pulsed cases.

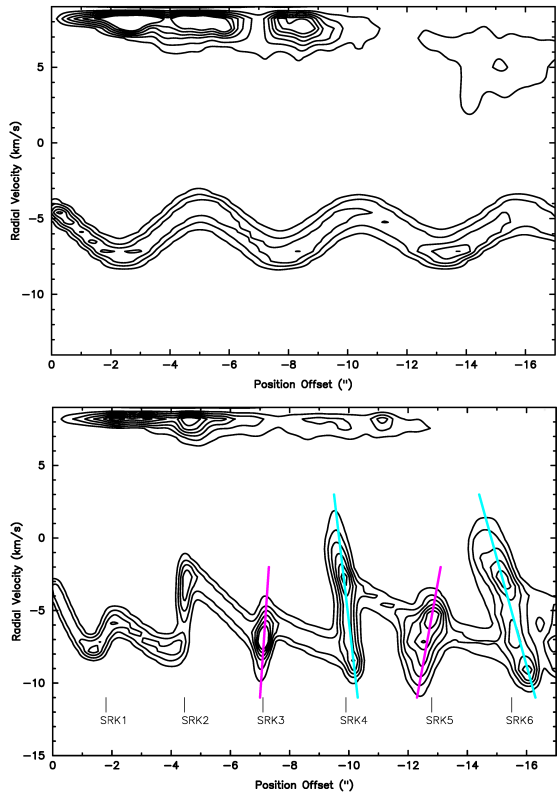

**Figure 4.** Position Velocity diagram along the jet beam for the steady velocity case (Top panel), and the 170±20 km s$^{-1}$ pulsed velocity case (Bottom panel). We focus only on the region of the red-shifted jet beam from position offset 0 to -17″, and radial velocity -15 to 9 km s$^{-1}$. The displayed contour levels begin at 0.1 Jy beam$^{-1}$ and step at intervals of 0.1 Jy beam$^{-1}$. This sensitivity was required to highlight the entire jet beam. We can see there is a distinct difference between the pulsed and steady velocity cases. Apparent velocity gradients may be detectable across the knots highlighted by the purple lines on SRK3 and SRK5, and the cyan lines on SRK4 and SRK6.

In the steady velocity case, the locus of the jet beam traces a regular sinusoidal wiggle-like pattern. The outermost visible contour covers a radial velocity range of ∼2.5 km s$^{-1}$ per period. This is consistent with the orbital motion velocity of 1.6 km s$^{-1}$ assuming the 5° viewing angle and velocity resolution effect of the PV diagram. Low velocity cocoon material is detectable near the systemic velocity at ∼9 km s$^{-1}$, but appears completely detached from the jet beam component at 0.1 Jy beam$^{-1}$ sensitivity.

In the 170±20 km s$^{-1}$ pulsed velocity case, the emission representing the jet beam material shows a structured pattern of shock jumps combined with the underlying wiggle-like pattern from the orbital motion. Unlike other observational PV diagrams of HH 211 such as Jhan & Lee (2015), we do not see an extended radial velocity range near the jet source. This is further evidence to support the idea of short period sub-knots that would lead to a more active PV environment near the source. Looking further from the source, the position of the knots are labelled SRK1–6. There are two competing effects; The physical size of the knots increase with distance from the source, while the magnitude of the radial velocity gradient increases. In addition, the direction of the gradient switches between each shock. The switch in gradient direction is highlighted in Figure 4 by the purple and cyan lines. Knots SRK4 and SRK6 in particular show a distinct measurable velocity gradient indicated by the cyan lines connecting the peaks of emission. In the PV diagrams of Jhan & Lee (2015), the authors also detect velocity gradients across the knots. They interpret them as the being associated with the forward and backward shocks within the knots. This is not the case in our model. Due to the wiggle motion and oblique shocks, the PV diagram does not cut entirely perpendicular through the knots. The apparent gradient in SRK4 and SRK6 is from left to right, and although less distinct, the gradient for knots SRK3 and SRK5 appears from right to left. When we compare the positions of these four knots to the analytical wiggle in Figure 2, we see SRK4 and SRK6 are on an upper edge of a wiggle, whereas SRK3 and SRK5 are on a lower edge. This suggests the apparent velocity gradient is solely related to the position of the knot along the wiggle. This can also be seen in Figure 1 where the orientation of the shock fronts are indicated by the blue lines in the lower panel. Each shock clearly has a different orientation, coincident with the orientations detectable in Figure 4. This shows that caution should be exercised if measuring radial velocity gradients from knots in PV diagrams as additional velocity components may be responsible.

### 3.5 Jet beam rotation

The direct detection of jet beam rotation is a popular topic at present. If jet beam rotation could be measured, then it would help constrain the physics of the actual jet launching mechanism. There have been several attempts at observational measurements of jet beam rotation (e.g. Lee et al. 2009; Choi et al. 2011), but as of yet, no conclusive results. The usual procedure to determine jet beam rotation is to measure a velocity gradient in a PV diagram cut perpendicular to the jet propagation direction. Theoretical models suggest jet beam rotation may have a toroidal velocity profile (Cerqueira et al. 2006) meaning a larger rotational velocity near the outer jet beam boundary and smaller rotational velocity near the jet beam center. Such components are not included in our model. Our model does not include an explicit jet beam rotation, only velocity components from the orbital motion at 1.6 km s$^{-1}$. As such, in each cut through the jet beam, the orbital velocity components all face the same direction. Either way, we present the PV diagrams of our current model to show how a non-rotating, but orbiting jet would appear.

In Figure 5 we display three PV diagrams cut perpendicular to be jet beam axis (into the page) through the first three fully developed knots along the jet beam; SRK2, SRK3 and SRK4. Contours are plotted at a higher sensitivity of 0.2 Jy beam$^{-1}$ with intervals of 0.2 Jy beam$^{-1}$. In each case, the jet beam component is clearly separated from the cavity wall component due to differing radial velocities between the fast moving jet beam and slower moving cocoon material. In the larger more developed knots (SRK3 and SRK4) the cavity wall component has an apparent reversed 'C' shape. This is due to absorption effects preventing us from detecting mate-rial at the backside of the cocoon. As we move downstream from SRK2 to SRK4, the intensity of the knots decrease and the apparent shape in PV space becomes more extended and elongated. SRK4 hints at being composed of two separate radial velocity components that may be clearer if the contours were shown at greater sensitivity. The elongated shape is due to the development of the oblique shocks mentioned in Section 3. In the downstream knots we are no longer looking purely perpendicular through the jet beam. This brief analysis shows the ideal location to measure a velocity gradient, if one existed, would be the most developed emission knots nearest the jet source. There, the emission would be strongest and measurements would be complicated the least by any additional velocity components from centrifugal effects.

### 3.6 HH 211 ALMA observation prediction

Whereas the SMA is capable of $0.3''$ resolution (e.g. Lee et al. 2010), the Atacama Large Millimeter Array (ALMA) Cycle-3 in 2015/2016 is capable of an unprecedented $0.05''$ resolution [2]. Here we create simulated CO intensity maps from our $170\pm20$ km s$^{-1}$ pulsed jet simulation convolved to $0.05''$ to make predictions for the feasibility of an ALMA HH 211 observation and the new science that it may reveal. The SMA convolved $0.3''$ image shown in Figure 5 can be compared directly to an ALMA convolved $0.05''$ image shown in Figure 6. At higher resolution, higher sensitivity is required. This time contour levels begin at 0.02 Jy beam$^{-1}$ km s$^{-1}$ with a step of 0.02 Jy beam$^{-1}$ km s$^{-1}$.

At $0.05''$ resolution the knots that were previously visible as extended emissions at $0.3''$ are resolved into sharper and more compact structures that more directly resemble the shape of the actual shocks seen in the hydrodynamic output from the PLUTO code in Figure 1. The more extended knots downstream begin to reveal some sub-structure. For example SBK6 indicated in Figure 6 where we can identify the peak of emission more accurately, but cannot detect the expected forward and backward shocks of the knot. This is likely due to the obliqueness of the the shock due to the orbital motion in our model, so it no longer resembles the classical picture of a shock. However, high-resolution high-sensitivity observations would be essential to determine if a knot is a single entity, as assumed in this paper, or composed of a collection of merged sub-knots, as hinted by the inability of our model to explain strong emission near the jet source.

The PV diagrams through the knots at $0.05''$ resolution are shown in the lower panels of Figure 6. Once again there is a separation between the low velocity cavity wall and the high velocity jet beam material that represents the knot. The PV structure has become narrower and sharper compared to $0.3''$ resolution. In this case, an even higher sensitivity of 0.005 Jy beam$^{-1}$ with a step of 0.005 Jy beam$^{-1}$ is required for the contours to adequately trace the detailed structure of the jet beam component. Compared to $0.3''$, the PV diagram of the jet beam component of SRK3 in particular now reveals two distinct peaks of emission. These may be components of the forward and backward shocks. It is less apparent in the more developed SRK4 knot, again due to the obliqueness and more extended size.

Overall, we see that an high-sensitivity, high-resolution ALMA observation of HH 211 should reveal a lot of new information about the structure of the knots and help us understand the true nature of the knots.

### 4 CONCLUSIONS

We study the impact of molecular chemistry on a 3D jet propagation simulation that is perturbed at its base through motion in a binary system. We reproduce the observed reflection symmetric wiggle of HH 211 using the analytical model of Lee et al. (2010) that considers the jet source to be moving as part of a protobinary system of period 43 years. This shows that orbital motion in a protostellar binary system can account for a reflection-symmetric wiggle structure.

Furthermore, we account for the observed knot pattern in HH 211 by using a sinusoidal velocity variation of period 21.5 years, exactly half the estimated orbital period. This assumption can reproduce the number of observed knots within the correct distance as in the SMA observation of Lee et al. (2010). However, it cannot precisely match the positions of all knots on each side, particularly the blueshifted side, suggesting an asymmetry to the velocity variation of each bipolar jet. Neither can our simple jet propagation model account for the near continuous emission near the jet source, or the the strength of the innermost knots, as seen in the HH 211 SMA observation. This is likely due to the lack of additional physics such as heating effects from an actual jet launching mechanism with a protostar and accretion disk. Alternatively, there could be a series of short period pulsations that lead to a series of closely spaced sub-knots creating the strong emission near the source, after which the sub-knots may later merge into the larger observed knots downstream.

We plot PV diagrams along the jet beam for both the pulsed and steady jet cases. We do not see an extended radial velocity range near the jet source, further evidence to support the idea of short period sub-knots. An interesting effect is seen downstream where gradients in radial velocity can be detected in the simulated knots. We show that they are a consequence of the wiggle motion, and not related to the forward and backward shocks.

Finally, we make predictions for an ALMA observation of HH 211 and show that its unprecedented angular resolution capability of $0.05''$ could resolve the knots into sharper shock structures that would help provide greater understanding of the physics of protostellar jets.

### 5 ACKNOWLEDGEMENTS

A. M, C.-F. L., and P. S. H acknowledge grants from the Ministry of Science and Technology (NSC 101-2119-M-001- 002-MY3, MOST 104-2119-M-001-015-MY3) and the Academia Sinica (Career Development Award) in Taiwan. B.V. would like to thank the support provided by Università degli Studi di Torino. Computations were performed on the computational facilities belonging to the Theoretical Institute for Advanced Research in Astrophysics (TIARA),

---

[2] http://www.almaobservatory.org/

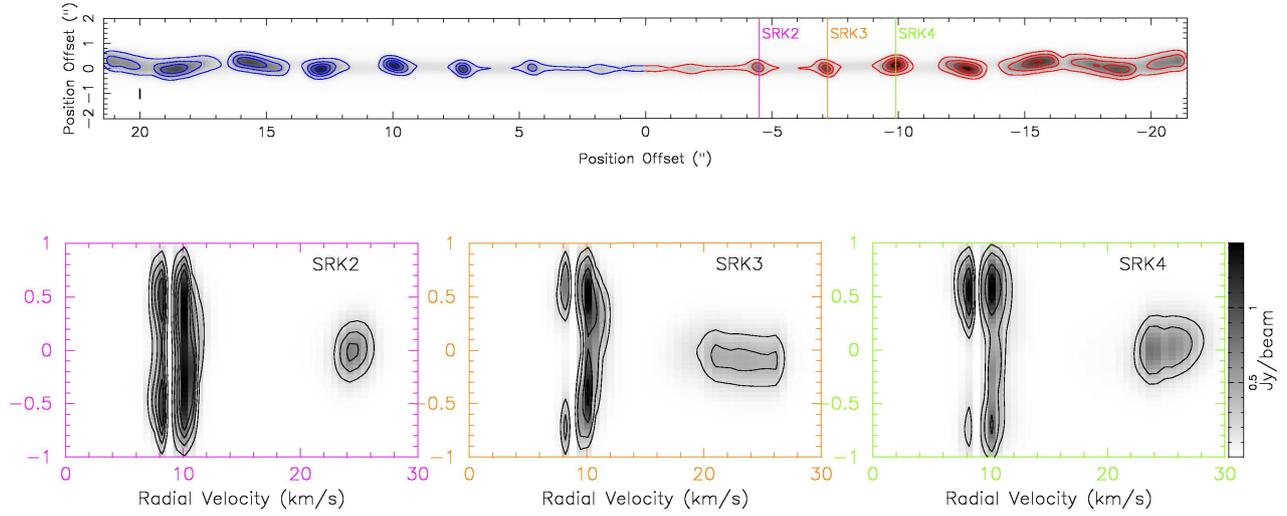

**Figure 5.** Integrated CO emission intensity map and position-velocity diagrams at three different positions through the redshifted knots SRK2, SRK3, and SRK5 whose positions are indicated by the colour coded vertical lines in the upper panel. The data is convolved for a resolution of $0.36'' \times 0.46''$ for direct comparison to the SMA observation. The upper CO emission channel map is the same as in Figure 2, i.e. the red and blue shifted components are integrated from 20 to 43 km s$^{-1}$ and -16 to 0 km s$^{-1}$, respectively, where the contours begin at 0.56 Jy beam$^{-1}$ km s$^{-1}$ and step at intervals of 0.84 Jy beam$^{-1}$ km s$^{-1}$. The lower PV diagrams cover the entire velocity channel range from -30 to 43 km s$^{-1}$ in order to highlight the distinction between the cocoon material component and the jet beam component. PV diagram contours are plotted at a higher sensitivity of 0.2 Jy beam$^{-1}$ with intervals of 0.2 Jy beam$^{-1}$. The PV diagrams show no velocity gradient as expected as our model only includes orbital motion. However, there are hints of some complexity in the older knots.

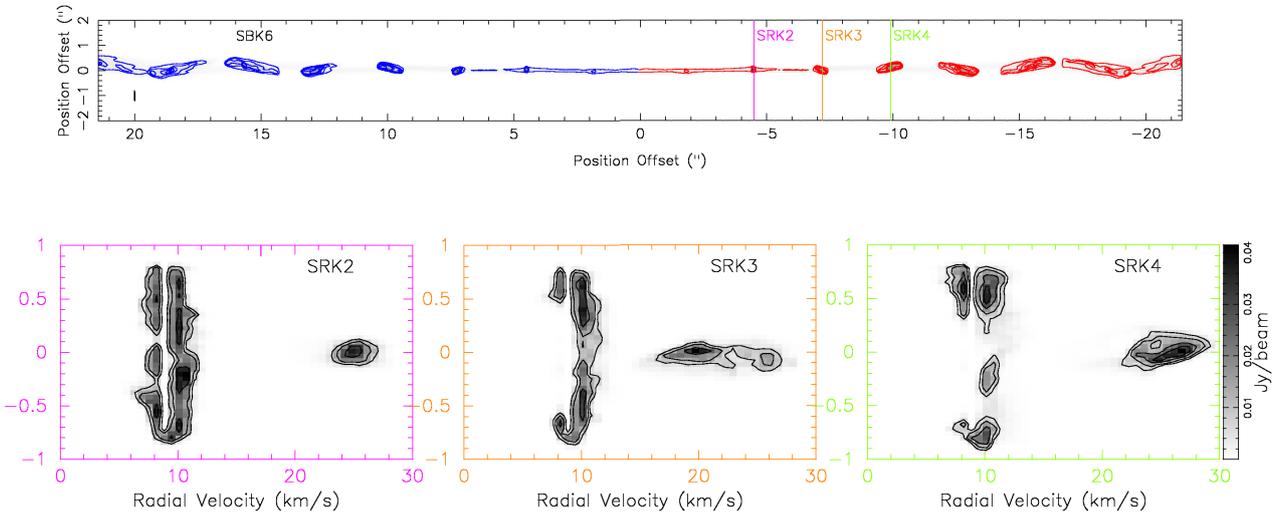

**Figure 6.** Integrated CO emission intensity map and position-velocity diagrams at three different positions through the redshifted knots SRK2, SRK3, and SRK5 as indicated by the colour coded vertical lines. The data is convolved for a resolution of $0.05''$ to represent the highest ALMA Cycle-3 capability. Extremely high sensitivity is required at this resolution, but ALMA is capable of it. The velocity range of the upper CO emission channel map is the same as in Figure 5, but the contours start at 0.02 Jy beam$^{-1}$ km s$^{-1}$ and step in intervals of 0.02 Jy beam$^{-1}$ km s$^{-1}$. Even at this level of sensitivity, the knots are only just detected. In the PV diagrams of the knots, contour levels of 0.005 Jy beam$^{-1}$ with intervals of 0.005 Jy beam$^{-1}$.

Academia Sinica, Taiwan. We are grateful to the referee whose comments improved the quality of the manuscript.